\documentclass{appolb}
\usepackage{epsfig}
\renewcommand{\thevolume}{34}
\renewcommand{\theyear}{2003}
\renewcommand{\theNo}{5}

\headtitle{Core-Collapse Supernova Mechanism -- Importance of Rotation}
\headauthor{A. Odrzywo\l{}ek et al.}
\begin{document}
\title{Core-Collapse Supernova Mechanism - Importance of Rotation}
\author{
A. Odrzywo\l{}ek$^b$
\thanks{
{\tt odrzywolek@th.if.uj.edu.pl}
}
M. Kutschera$^{a,b}$
M. Misiaszek$^b$\\ and
K. Grotowski$^{a,b}$
\address{$~^{a)}$H. Niewodnicza\'{n}ski Institute of Nuclear Physics\\ Radzikowskiego 152, 31-342 Krak\'{o}w,
Poland\\
$~^{b)}$M. Smoluchowski Institute of Physics, Jagellonian University\\ Reymonta 4, 30-059 Krak\'{o}w,
Poland} }
\maketitle
{\it  (Received July 17, 2002)}

\begin{abstract}
An attempt is made to assess the significance of rotation in the
core-collapse supernova phenomenon, from both observational and theoretical
point of view. The data on supernovae particularly indicative
of the role of rotation in the collapse-triggered explosion is emphasized. The
problem of including the rotation of presupernova core into the supernova
theory is considered. A two-dimensional classification scheme
of core-collapse supernovae is proposed which unifies ``classical'' supernovae
of type Ib/c and type {\sc II}, ``hypernovae'' and some GRB events.

\end{abstract}

\section{Introduction}

{

The phenomenon of supernova amazed already ancient observes as some bright
historical supernovae were visible on the sky even in the daytime. The
absolute luminosity of supernovae has been properly estimated
only in the 20-th century, when it was realized that supernovae belong to a
very
special class of astronomical events. In 1885 the nova S~And appeared in
the M31 nebula, now well-known as the Large Galaxy in Andromeda. At those
times many astronomers accepted the in-Galaxy theory of M31 and other nebulae.
After establishing that the real location of M31 is extragalactic, astronomers
were
forced to conclude the nova S~And\footnote{Now called SN 1885A.} was much
brighter than any usual nova \cite{Trimble} - it was a
\mbox{super-nova!}

The systematic supernova research began in the 20-th century. Unfortunately,
there was no Galactic supernova event since the 17-th century. In spite of this
astronomers have observed more than 2000 extragalactic supernovae. The number
of observed events grows rapidly, from about 20 per year in the eighties to
about 200 per year now. In contrast to the optical events, more than 600
supernova remnants have been found in the Galaxy. Also, a number of
extragalactic remnants, mainly in the Local Group galaxies LMC, SMC, M31 and M33
\cite{WeilerSramek} have been observed.

In 1942 Minkowski \cite{Minkowski} introduced the modern classification scheme
of supernova events into two classes. To the first class belong supernovae with
no hydrogen absorption lines in the spectrum referred to as type
{\sc I}~supernovae. The second class comprises the supernova events with strong
hydrogen lines which are referred to as type {\sc II} supernovae.

As for the physical nature of supernovae, Landau \cite{Landau} in 1932, soon
after discovery of the neutron, suggested the possibility of existence of dense
stars composed of neutrons which are stabilized by very high pressure of the
neutron gas. In 1939 Baade\&Zwicky \cite{Zwicky&Baade} proposed the
gravitational collapse of a normal star to such a neutron star as the
supernova energy source. This picture is generally accepted today. In 1960
Fowler and Hoyle \cite{F&H,H&F} pointed out that also nuclear reactions can
serve as a source of the supernova energy. They proposed the thermonuclear
explosion of a white dwarf or a giant star as an another possible supernova
mechanism.

After several decades that passed since the original proposal of Zwicky and
Baade, significant progress in understanding the supernova mechanism has been
achieved. The detection of the neutrino burst correlated with the appearance of
SN1987A proved that the theoretical research is on a right track. Unfortunately,
the standard core-collapse supernova theory suffers from difficulties in
producing a successful explosion under general conditions. This could be a
consequence of suppressing star's rotation in the theory. In this paper we
collect arguments in favour of the necessity to include rotation into the
supernova theory. The paper is organized as follows. In the next section we
review in some details the modern version of supernova classification. We wish
to emphasize a division of observational data into these related to outer layers
of the exploding star and the ones reflecting the physics of the engine
mechanism. In Sect.\ref{StdSN} the essential features of the standard supernova
theory are reminded. The problem of inclusion of the presupernova core rotation
into the supernova theory is considered in Sect.\ref{Rotation}. Finally, in
Sect.\ref{LastSec}, we introduce a two-dimensional classification employing some
measure of the rotation as a second dimension.

}

\section{Modern classification of supernovae}

{

With a simple classification of supernova events available, researchers were
tempted to establish a direct relationship between the two proposed
supernova mechanisms and the two observed supernova types. Unfortunately, the
relation of the thermonuclear explosion mechanism and the core-collapse
scenario to type {\sc I} and type {\sc II} supernovae is not as straightforward
as it may seem. It turns out that type {\sc I} supernovae with no hydrogen lines
form a class of diverse events which can be further divided into more
homogeneous subclasses. The spectra near the maximum brightness are used to
distinguish type {\sc I}a events which show a strong SiII absorption dip at
$\lambda\; 6150\; {\rm \AA}$ and type {\sc I}b/c events with no or weak SiII lines.
Further, a strong helium line at $\lambda\; 5876\; {\mathrm \AA}$ is employed to
distinguish type {\sc I}b events from those of type {\sc I}c. Type {\sc I}b/c
supernovae occur in the same environment as do type {\sc II} supernovae, namely
in the star forming regions in spiral galaxies. Both are thus related to young
population {\sc I} stars. Type {\sc I}a supernovae are more common and they
occur in all types of elliptical and spiral galaxies, and in the halo of our
Milky Way galaxy. This location indicates they are related to population {\sc
II} stars. The main observational features used as a basis of modern supernova
classification are presented in Table I .

\begin{table}[!t]
\begin{center}
{\sc Table I}
\end{center}
{Supernovae classification and properties.}
\begin{center}
\includegraphics[width=\textwidth]{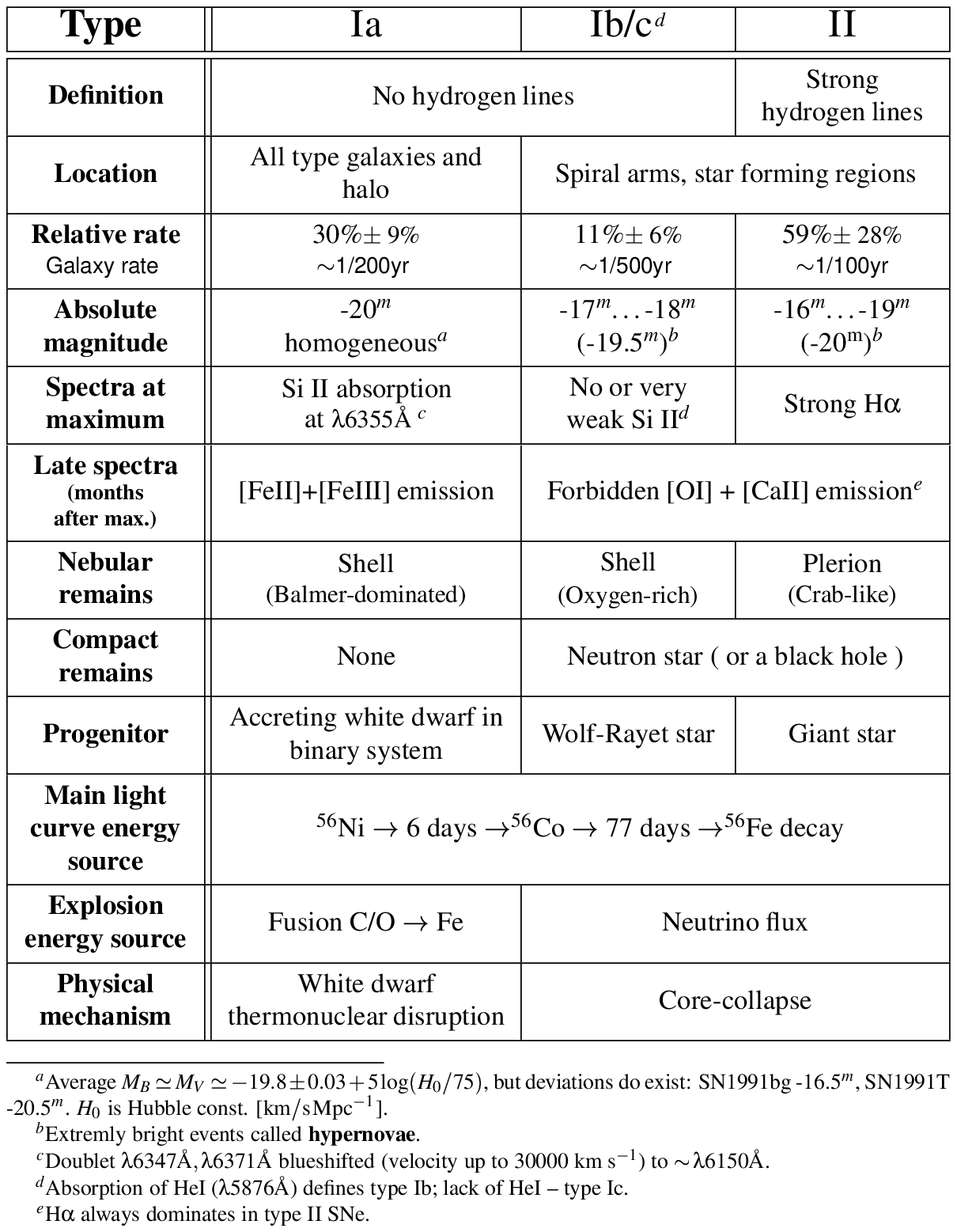}
\end{center}
\end{table}

A generally accepted hypothesis with regard to the relationship of the physical
supernova mechanism and the observed supernova types postulates that
thermonuclear explosions of white dwarfs produce type {\sc I}a supernovae
and the core collapse of giant stars results in type {\sc I}b/c/{\sc II}
supernovae. Basic arguments in favour of this identification come from
common locations of type {\sc I}a supernovae in old stellar systems and type
{\sc I}b/c/{\sc II} - in regions of active stellar formation in spiral galaxies.

Homogeneity of type {\sc I}a supernovae also indicates the same explosion
mechanism. Type {\sc I}a events have similar spectral properties and obey a
simple empirical relation (\eg the Phillips relation \cite{Phillips}) between
the maximum absolute brightness and the behavior of the light curve, which
states that brighter events are longer. The observed scattering of absolute
magnitudes at the maximum is about $2^m$, and further division of type {\sc I}a
into subclasses is not excluded \cite{Leibundgut}.

It was long believed that these differences are due to observational errors and
all type {\sc I}a supernovae are almost identical. However, the maximum
brightness and the decay time are possibly related to the amount of $^{56}$Ni
synthesized during the explosion. The light curves clearly show that the beta
decay of $^{56}$Ni into $^{56}$Co, and a subsequent decay of $^{56}$Co to
$^{56}$Fe with a half-life of 77 days, are the main energy sources supporting
the light emission during the late phase of the explosion. It is of great
interest that the radioactive energy source from $^{56}$Ni and $^{56}$Co decays
occurs in all types of supernovae in spite of very different explosion
mechanisms.

Significant constraints on the supernova theory are imposed by observations of
the nebular remnants. Unfortunately, in spite of known positions of SN1006,
SN1054, SN1572 and SN1604 we can not firmly classify these historical
supernovae. In effect, it is not perfectly clear what remains after type {\sc
I}a/b/c and {\sc II} supernova explosions. We can distinguish three basic types
of remnants:

\begin{itemize}
\item Balmer-dominated shell
\item oxygen-rich shell
\item plerion (Crab-like nebula)
\end{itemize}

A classical example of a plerion is M1, the Crab nebula. Because the pulsar is
found in its center we connect plerions to type {\sc II} events. They are
sometimes surrounded by weak shells. The Balmer-dominated shell is usually
connected to type {\sc I}a, and its emission can be explained as a result of the
interaction between the shock wave and the interstellar hydrogen. Arguments
confirming this identification come from the analysis of SN1572 and SN1604
remnants, which both probably were type {\sc I}a supernovae. In addition, we can
not see any compact stellar remnant, such as a pulsar or a neutron star, inside
or near outside the shell. Finally, type {\sc I}b/c supernovae are supposed to
leave as a nebular remnant the oxygen-rich shell. This identification is
supported by the theory of origin of type {\sc I}b/c supernovae from stars
stripped off hydrogen and helium layers. The lost matter at least partially
fills in the space around donor stars. If extensive mixing and core dredge-up
took place at some evolution phases of these stars, we expect some enrichment of
circumstellar medium (CSM) in heavy elements. The shock interacts with CSM whose
composition is a relic of star's earlier evolution and excites atoms of heavier
elements, like \eg oxygen.

Observed spectral differences between various types of core-collapse supernovae
are in significant part a result of the differences in the outer shells.
Classification of core-collapse events clarifies and becomes more elegant when
we use the lost mass as a basic ordering parameter \cite{CappellaroTuratto}.
Stars with very massive hydrogen layers at the onset of the explosion produce
type {\sc II}-P supernovae with the light curve plateau. An extreme example of
such a light curve was provided by the SN1987A. If the star has not very massive
hydrogen shell the explosion is classified as type {\sc II}-L with fast, linear
decay of the light curve. In case of a very thin H shell the explosion is type
{\sc II}b and if the whole hydrogen is lost, the event is classified as {\sc
I}b. If in addition the helium layer is also removed we can see type {\sc I}c
supernova. This ordering of core-collapse supernovae is displayed schematically
in Fig.~\ref{1D}.

\begin{figure}
\begin{center}
\includegraphics[width=\textwidth]{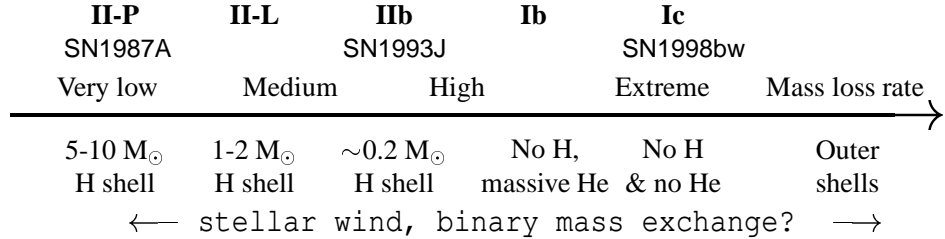}
\end{center}
\caption{Unification of the core-collapse events within an one-dimensional
family of the growing mass loss rate.
}
\label{1D}
\end{figure}

Within this scheme, every type of core-collapse supernovae covers a range of
outer shell masses of pre-supernova stars. For every pair of supernovae we may
find a third one which could be placed between them in the classification
scheme. For example, SN1993J in M81 was a missing link between type {\sc II} and
type {\sc I}b with a residual hydrogen shell, now distinguished as type
{\sc II}b.  The evidence of supernovae corresponding to stars with thin He
layers is not clear yet, but candidates exist \cite{Filippenko}. The fact this
classification works indicates that mass loss rates in progenitors increase from
type {\sc II} to type {\sc I}c. Outer layers are expelled by strong stellar
winds, but in extreme cases also stripping due to the mass exchange in close
binaries occurs, as we indicate in Fig.~\ref{1D}.

This scheme clearly shows that the engine mechanism is not of major importance
for the classification of supernovae based on light curves and spectra, which
both are formed by processes occurring in outer layers of the exploding star.
There are, however, a number of observations related to supernovae, such as
pulsar initial velocities, asymmetry of explosions, ``hypernovae'', jets and
possibly some gamma-ray bursts (GRB), which are likely related to processes in
the inner part of the exploding star, in particular they can directly reflect
physical conditions prevailing in the collapsing core. The detailed picture of
the collapse depends on the mass of the core related to the ZAMS mass and the
evolutionary track of the progenitor. In this paper we focus on the rotation of
the progenitor core which can very significantly affect physics of the core
collapse. Below, in sect.\ref{LastSec}, we discuss interpretations of the above
observations as an evidence of rotation of the collapsing core. We propose a
two-dimensional classification of the core-collapse supernovae which can
accommodate "hypernovae" and GRB's. It is an extension of the scheme shown in
Fig.~{\ref{1D}, with the second dimension being some measure of the amount of
rotation of the collapsing core, Fig.~\ref{2D} on page~\pageref{2D}.

}

\section{Standard Supernova Mechanism \label{StdSN}}

{
In the standard scenario of the core-collapse supernova without rotation, a
massive star at the onset of collapse is a red or blue super-giant, for example,
the progenitor of SN1987A was a B3I spectral type star. It contains a number of
shells, which correspond to the following stages of nuclear burning, given here
in a simplified form:

\begin{tabular}{lcl}
{~~H}&$\longrightarrow$&{$^4$He}
\\
{~$^4$He}&$\longrightarrow$\hspace{5mm}&{$^{12}$C}, {$^{16}$O}
\\
{$^{12}$C} ( {$^{16}$O} )&$\longrightarrow$&{$^{20}$Ne}, {$^{24}$Mg}
\\
{$^{20}$Ne} ( {$^{16}$O}, {$^{24}$Mg} ) &$\longrightarrow$&{$^{16}$O},
{$^{24}$Mg}
\\
{$^{16}$O} ( {$^{24}$Mg}, {$^{28}$Si} )
&$\longrightarrow$&{$^{28}$Si}
\\
$^{28}$Si ( $^{32}$S ) &$\longrightarrow$& $^{56}$Ni, $^{56}$Fe, $^{54}$Fe
\end{tabular}

The synthesis of the most strongly bound nuclei near $^{56}$Fe ends the network
of thermonuclear reactions. In Table II we show main evolutionary phases of a
star of 25$M_{\odot}$. One can notice that burning of neon begins about one
year before the star's death.

\begin{table}
\begin{center}
{\sc Table II\\}
\end{center}
{ The evolution of $25M_{\odot}$, $Z=0.02$ star according to
\cite{Chieffi}. Duration of a given phase is defined here as time since central
ignition of a given reaction until the central ignition of a next reaction. This
includes phases of off-center burning, with no reactions in the center. The
temperature and density correspond to the beginning of each phase.
}
\begin{center}
\includegraphics{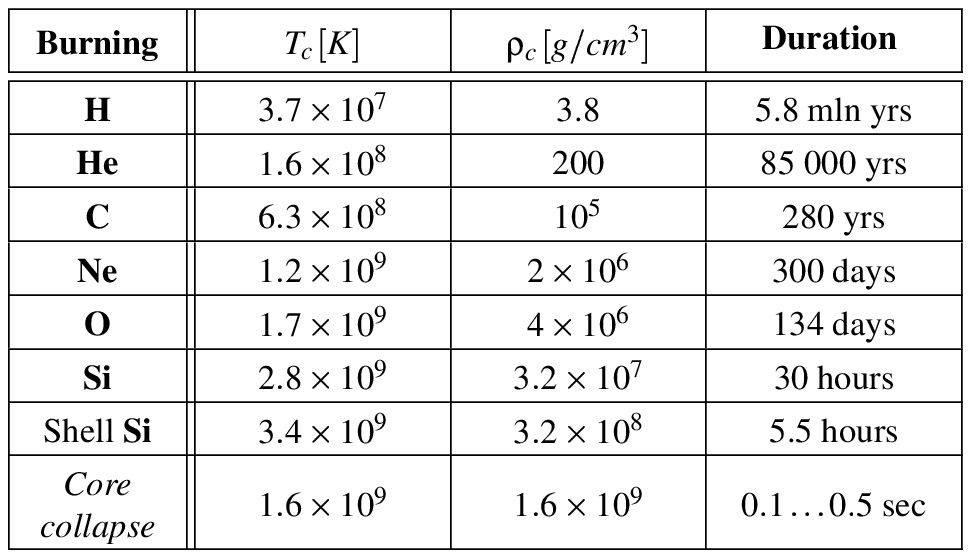}
\end{center}
\end{table}

The late phases of nuclear burning are quite
complicated and still not perfectly understood. Soon after ignition of
an off-central silicon burning the core composed of iron-group nuclei loses its
stability and starts to collapse. A short explanation is that the core mass
exceeds the Chandrasekhar limit. The full expression of the Chandrasekhar mass
is \cite{Bethe}:
\begin{equation}
{M_{Ch}}\,=\,1.44\,M_{\odot}\,
{(2\,Y_{e})}^2\,
\left[ 1 + {\left( \frac {S_e}{\pi Y_e}
\right)}^2 \right]\,
\left[1-\frac {3}{5}\,{\left(\frac
{12}{11}\right)}^
{1/3}
\,\alpha\,{\bar {Z}
}^
{2/3}
+\frac{p_{rad}}{p_{mat}}\right].
\end{equation}
It is a function of the electron fraction $Y_e$, the entropy per baryon $S_e$,
the nuclear composition expressed through an average nuclear charge $\bar{Z}$,
and the ratio of radiation and matter pressure $p_{rad}/p_{mat}$; $\alpha
\simeq 1/137$ is the fine structure constant. For $Y_e=0.5$, $S_e=0$ and
$p_{rad}=0$ we obtain the famous result $M_{Ch}=1.44\,M_{\odot}$. The mass of
the stellar ``Fe'' core is in the range \mbox{1.2 - 2 $M_{\odot}$}. A number of
processes are important in the last day before the collapse. The electron
capture by iron and nickel isotopes decreases the electron fraction in processes
such as \eg:\\
\begin{displaymath}
^{55}\textrm{Fe}+e^-\longrightarrow^{56}\textrm{Mn}+\nu_{e}.
\end{displaymath}
Neutrinos carry away the energy from the core decreasing its entropy. The
silicon burning Si $\rightarrow$ Fe stops in the core and begins in a
surrounding shell. The most important reason of the collapse onset is the
decrease of entropy. Usually the Chandrasekhar mass limit is not really exceeded
because of a huge external pressure not included in $M_{Ch}$. The instant when
the core starts to collapse is deduced from the evolutionary code results.
When the speed of contracting matter exceeds some in advance prescribed value,
the evolutionary track is finished. This moment is believed to be close to the
real stability loss and to start of actual collapse of the core.

Once started, the collapse is fast, close to a free fall. In the numerical
simulations the collapse lasts 100 - 500 milliseconds, depending on the initial
conditions and on the input physics. During the first phase the collapse is
homologous -- the velocity is proportional to the radius. The sound speed
decreases with the radius, and at a sufficiently high distance from the center
it falls below the matter speed. This place is called the sonic point. Time
when the central density reaches its maximum defines the core bounce. At this
time the infalling matter is stopped by the pressure component due to nuclear
forces which starts to grow very rapidly. In milliseconds the velocity goes to
zero and simultaneously the density grows by a few orders of magnitude
\cite{Burrows}. Strong sound waves start to propagate outside. As a result of
non-linear hydrodynamical effects and the supersonic speed of matter
discontinuity forms in velocity, density, pressure and entropy -- the sound
wave becomes the shock wave.

It was accepted for a long time that the sketched above process is a main part
of the real explosion mechanism and is able to provide at least a few
foe\footnote{foe -- fifty-one-erg -- $10^{51} \,$erg}  of energy. The shock was
believed to traverse the entire star and reach the surface to produce a
spectacular supernova type {\sc II} (or~{\sc I}b/c) event. This scenario is
known as a prompt explosion. Unfortunately, more detailed calculations with
improved physical ingredients, such as a realistic equation of state (EOS),
general relativity corrections, sophisticated progenitor models and a complete
set of nuclear reactions included in new gas-dynamical algorithms, have shown
the failure of this idea. The shock wave gets stuck in the Fe core as a result
of the energy loss chiefly due to heavy nuclei dissociation
($\sim8.7\,$MeV/nucleon $\simeq1.7$foe$/0.1M_{\odot}$) and -- when shock moves
out of the neutrino trapping sphere -- the neutrino emission. Although it is
still possible to produce a successful prompt shock under extreme assumptions on
the equation of state or other part of included physics, nowadays neutrino
processes are believed to play a major role in core-collapse supernova
explosions, so let's look at them.

As we pointed out already, beyond the nuclear matter density, the EOS
stiffens very rapidly. Simultaneously matter initially composed of nuclei and
electrons, transforms into a nucleon-electron gas and finally becomes an almost
pure neutron matter. Other phases of dense matter, such as \eg the
kaon-condensed nuclear matter or the quark-gluon plasma can possibly form, but
usually are not considered in the standard supernova scenario. Almost every
electron-proton pair is transformed into neutron and neutrino. Initially
neutrinos escape freely. But as a result of neutrino cross-sections growth with
temperature and density, neutrinos become trapped. It means that the diffusion
time is bigger than the dynamical time-scale. The gravitational energy released
in the collapse is ``frozen'' as the energy of the lepton Fermi sea. The inner
part of the star which is not blown off, after the core bounce and shock
traversal becomes a protoneutron star (PNS). The edge of the neutrino-trapped
zone is referred to as the neutrinosphere, with analogy to a photosphere of
normal stars. Definition of the neutrinosphere is somewhat ambiguous as
different neutrino flavors have different and diffused neutrinospheres. Harder
neutrinos escape from bigger radius than the do soft ones, because of the
cross-section dependence on the neutrino energy. The most important fact is the
location of the neutrinosphere between the shock and the neutrino-rich
protoneutron star. The trapped neutrinos diffuse out of the protoneutron star in
time of the order of one second, and simultaneously the protoneutron star
contracts from an initial radius of about 60km to final radius of 20km, like a
Kelvin-Helmholtz star \cite{KeilJankaMuller}. Neutrinos carry away about
100~foe of energy, \ie $\sim$99\% of the total energy released in the collapse.
The neutrino flux vanishes in tens of seconds. This picture was in general
confirmed by the detection of neutrinos during the SN1987A event and now this is
the most well established part of core-collapse supernova theory. In spite of
very small neutrino cross-sections about 1\% of the energy is transferred to the
hot radiation bubble between the nascent neutron star and the shock in hundred
milliseconds. It causes the shock wave revival. Later, the explosion is similar
to that of a prompt mechanism. This scenario is called a delayed or
neutrino-driven supernova mechanism. Currently it is a subject of very active
studies. The neutrino transport and convection appear to be the most important
processes of this scenario \cite{Janka}.

The shock produced in the engine of supernova by prompt or delayed mechanism
traverses the entire star ionizing the gas, igniting nuclear reactions and
triggering convection \cite{Kifonidis1,Kifonidis2}. The shock wave may reflect
at the shell boundaries. The onion-like structure of different nuclei layers
is destroyed. Intensively mixed matter from the center may move close to the
surface. In a few hours the shock reaches the surface. The photosphere begins to
expand and we can see the enormous growth of brightness. A supernova appears on
the sky. During the first few weeks the light curve is dominated by the
recombination of ionized atoms. Later, the light curve mimics the decay curve of
beta-radioactive $^{56}$Ni. The subsequent decay of $^{56}$Co with $T_{1/2}=77$
days is the energy source for supernova during next months.

}

\section{Rotation in Supernova Theory \label{Rotation}}

A major deficiency of the sketched above supernova theory is suppression of
rotation of the presupernova star. From the observational point of view there is
growing evidence that rotation can play an important role in the explosion.
Below we discuss some relevant data. However, before we attempt to asses if
inclusion of rotation is necessary to explain the observed properties, we
address the following important questions:
\begin{itemize}
\item are massive stars rapid rotators?
\item do cores of massive star rotate?
\item what is the shape of the rotating core?
\item \label{Sec.3} what is caused by the angular momentum
conservation during collapse?
\end{itemize}

\subsection{The simplest rotating star}

{

Let us consider the behavior of a rigidly rotating body which is incompressible
and homogeneous and bound by the Newtonian gravity. This is very simple and
idealized model of a rotating star, closer to the liquid drop model of the
atomic nucleus \cite{Swiatecki} except of the surface tension, which is
negligible for astrophysical objects and huge for atomic nuclei.\footnote{In a
ball of water the gravitational
energy exceeds the
surface tension energy if radius is bigger
than 10 m.}
It illustrates, however, basic features of the rotation's influence on
properties of real objects.

The most important parameter, which determines physics of rotating and
gravitating bodies, is the ratio of rotational to gravitational energy,
$E_{rot}/E_{grav}$ (here $E_{grav}$ is the absolute value of the gravitational
energy). In case of zero angular momentum the body is of spherical
shape. Any rotation leads to a spheroidal shape, as shown by Mac Lauren in
1742. But when $E_{rot}/E_{grav}$ exceeds the value of 0.1375 two solutions of
the problem exist (Jacobi, 1834). The first solution is the Mac Lauren spheroid,
and the second one is a triaxial ellipsoid. The latter is the ground state
corresponding to the minimum of the sum of rotational and gravitational energy.
Transition from the Mac Lauren spheroid to the Jacobi ellipsoid requires some
dissipation mechanism, because of difference in the total macroscopic energy.
Mac Lauren spheroids beyond $E_{rot}/E_{grav}=0.1375$ are secularly unstable
with respect to dissipative processes, such as, \eg viscosity.

When the rotational frequency of the body increases, for example in effect of
shrinking, two cases are possible. If the dissipative time-scale is short
compared to the dynamical time-scale, the body evolves through a sequence of
Jacobi ellipsoids. For $E_{rot}/E_{grav}> 0.16$ they are unstable secularly and
form so-called Poincare pears. A bit more rotational energy leads to the
dynamical instability and to the fission. If dissipative processes are slow the
body evolves along a sequence of Mac Lauren spheroids, which are dynamically
stable up to $E_{rot}/E_{grav}=0.2738$. Beyond this limit no stable
configuration exists and the body has to get rid of the angular momentum or to
breakup.

Surprisingly, inclusion of a differential rotation and a compressible equation
of state (and also general relativity corrections) results in minor changes of
the secular and dynamical instability limits of $E_{rot}/E_{grav}$ which are
important for supernova theory. Recent publications report the possibility of
significant decrease of those limits, for a toroidal density stratification. The
secular instability limit of the ratio $E_{rot}/E_{grav}$ becomes 0.038
\cite{Imamura}, for transitions leading to bar-like configurations. The
dynamical instability limit is found to be $E_{rot}/E_{grav}= 0.14$
\cite{Centrella}.

During the collapse of the stellar core the ratio $E_{rot}/E_{grav}$ grows
significantly, by a factor of a few tens. Given the insensitivity of the
instability limits with respect to details of rotation and to the
compressibility of matter, the values listed above seem to be sufficiently
accurate to allow us to assess the possibility of triaxial deformations and the
core breakup in SN events.

}

\subsection{ Rotation of supernovae progenitors}

{

Are supernova progenitors rapid rotators? The answer is well known to
astronomers \cite{Fukuda}. Single stars which are supposed to be supernova
progenitors begin their lives as O and B main sequence stars with initial masses
$M>8M_{\odot}$. The surface velocity is very high for these stars. It can be
determined observationally with some uncertainty due to an unknown angle between
the rotation axis and the observer direction. The velocity may be close
($\sim$70\%) to keplerian velocity at the surface radius. But, except of the
Sun\footnote{Helioseismology allows us to
see the interior of
the Sun. Other stars are in the range of observational
abilities too; see \eg. \cite{Carrier}.},
it is impossible to determine rotation inside the star. To address the problem
one must resort to stellar modeling.

In last few years new detailed evolutionary calculations of rotating stars have
been carried out~\cite{HegerandLanger}. The results are in good agreement with
observed surface properties. One can thus treat predictions with regard to
the internal structure with some confidence. Contrary to previous opinions,
numerical results show that cores of stars rotate fast with velocity not
significantly dependent on the initial conditions. Nevertheless, one should
mention that the calculations described above neglect magnetic fields which may
transport angular momentum and slow down the core.

Typically, according to \cite{HegerandLanger} Fe cores have radii of about 2000
km, masses of $1.5M_{\odot}$ and rotate differentially with periods of the order
of 10 seconds. Outer convective shells surrounding cores rotate rigidly, and the
rotational frequency drops at shell boundaries discontinuously. The ratio
$E_{rot}/E_{grav}$ reaches near the center the maximum values up to 0.04. It was
assumed in calculations of \cite{HegerandLanger} that it is safe to neglect
the possibility of non-axisymmetric core deformations because
$E_{rot}/E_{grav}=0.04$ is only $\sim$30\% of the Mac Laurin spheroid secular
instability limit. Results cited in the previous subsection \cite{Imamura}
suggest the secular instability for toroidal ``fizzler"\footnote{This is a
transient object between those stabilized by the degenerate electron gas
pressure (\eg ``Fe'' cores and white dwarfs) and neutron stars. Its existence
and dynamical stability are results of extremely strong centrifugal force, but
``fizzlers'' are secularly unstable.} configurations at 0.038 and the
possibility of ``triaxial'' cores during last phases of the stellar evolution
seems not to be excluded if the core is far from a spheroidal shape. In spite of
the rapid core rotation almost all angular momentum is located in distant
massive shells, but $E_{rot}/E_{grav}$ is very small there, as indicated in
Table~III.
\begin{table}
\begin{center}
{\sc Table III\\}
\end{center}
{Parameters of rotation of inner shells of a typical presupernova star
from~\cite{HegerandLanger}.
}
\begin{center}
\includegraphics{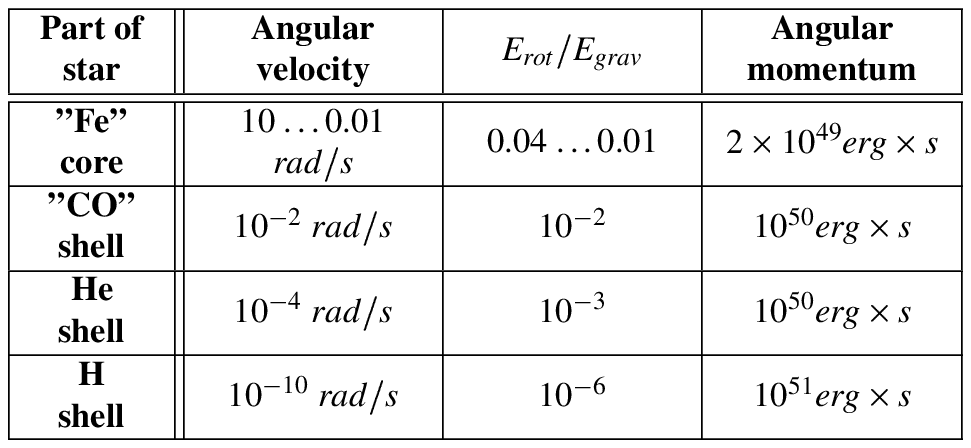}
\end{center}
\end{table}

\subsection{The shape of rotating star}

{

Calculations presented above are based on the assumption that the density
stratification is not far from spheroidal and the rotation is close to
``shellular", \ie the angular frequency is constant on a family of spheroidal
surfaces. The set of stellar structure equations is modified due to rotation but
this method preserves the conventional description of the stellar structure in
terms of a one-dimensional mass coordinate, with constant mass surfaces
being non-spherical \cite{HegerandLanger}. In this approach some processes,
such as the transport of angular momentum, the mixing of stellar matter and
nuclear reactions are well described. However, the real two-dimensional
structure is treated only approximately.

\begin{center}
\begin{figure}
\centering
\includegraphics[height=0.9\textwidth,angle=270]{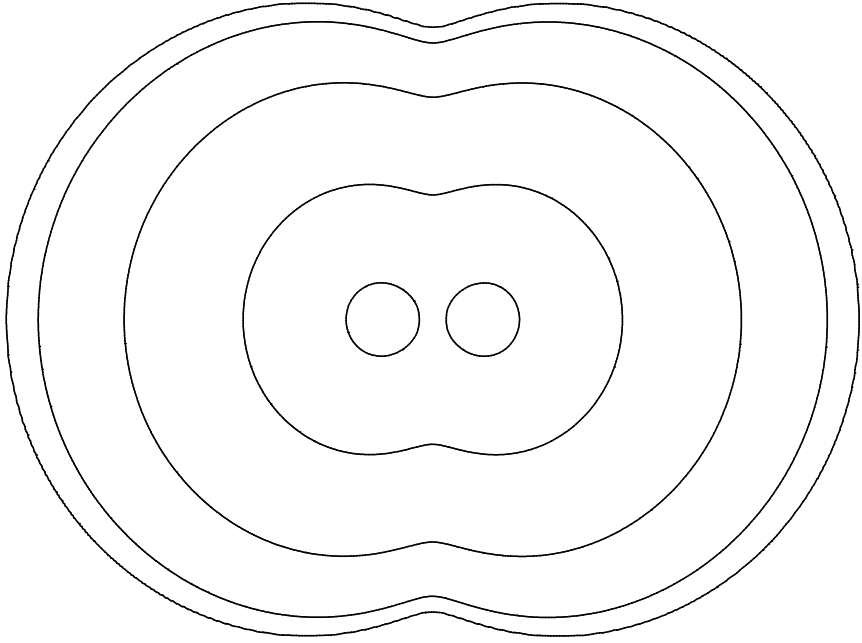}
\caption{An example of the iso-density contours for
n=3/2 ($p=K \rho^{5/3}$  EOS) polytropic model with a cylindrical distribution
of the angular frequency, \mbox{$\Omega(r)=\Omega_0/(1+r/A)$},
concentrated near the rotation axis
($A=0.1\,R_0$, where $R_0$ is the equatorial radius)
calculated using eq.~(\ref{firstorder}).  The off-center density maximum, the
flattened shape, the toroidal density distribution in the central
region and the
``cusp'' in the polar region are common features
of rapidly and differentially rotating polytropes.
}
\label{Isodensity}
\end{figure}
\end{center}

Such an approach is especially unsatisfactory when we attempt to analyze the
collapse of the core because the star model {\em is not in full mechanical
equilibrium}. The situation is analogous to starting the simulations of the
rotating core collapse from the initial configuration which is that of
a non-rotating core endowed with some amount of rotation.
This case has been analyzed in \cite{ZwergerandMuller}, where it was
shown that some of models without initial equilibrium evolve in a significantly
different way, compared to initially hydro-stationary models. In this way only
rough estimates of the basic quantities, such as \eg   the density at the core
bounce, can be obtained. The analysis has been performed in the axisymmetric
case with the cylindrical rotation and the barotropic EOS. One may expect a
similar result for an arbitrary rotation law of the initial model.

An important conclusion from the above considerations is that the exact
density distribution and the velocity field inside the presupernova star in the
full hydro-stationary equilibrium are required as an input for the analysis of
the collapse of the core and its later evolution. To find the exact density
distribution, in particular the shape of the stellar surface and the iso-density
contours inside the star, one has to solve the equations of hydrostatic,
or more explicitly hydro-stationary, equilibrium, taking into account the
non-zero velocity field. This has been achieved in very simplified stellar
models only. For the equation of state \mbox{$p=p(\rho)$}, including
polytropes, the rotation was assumed to be cylindrical with angular velocity
being a function of the radial variable in cylindrical coordinates only
\cite{EriguchiMuller}. Under these conditions one can derive an
analytic first-order approximation to the resulting integral equation for
the axisymmetric density distribution:
\begin{equation}
\label{firstorder}
\rho_1(r,z) = f^{-1} \left [ \, f(\rho_0) -
\Phi_c(r) + \left < {\Phi_c} \right > \right ],
\end{equation}
where $\rho$ is the density as a function of cylindrical coordinates
$(r,\phi,z)$, and $\left < {\Phi_c} \right >$ is the centrifugal potential
$\Phi_c$ corresponding to the angular frequency distribution inside the star
$\Omega(r)$,
\begin{equation}
\Phi_c(r) = \int_0^r \tilde{r}
\Omega(\tilde{r})^2 d\tilde{r},
\end{equation}
averaged over the volume $V_0$ of a non-rotating star with the density
$\rho_0(r)$ and the radius $R_0$:
\begin{equation}
\left < {\Phi_c} \right > = (\frac{4}{3} \pi
{R_0}^3)^{-1} \int_{V_0} \Phi_c(r)\, dV.
\end{equation}
In eq.~(\ref{firstorder}) $f(\rho)$ is
the enthalpy\footnote{The density in eq.(2) is the inverse function $f^{-1}$:
$f(f^{-1}(\zeta)) = f^{-1}(f(\zeta)) = \zeta$.}
of the barotropic gas with an EOS $p=p(\rho)$:
\begin{equation}
f(\rho) = \int \frac{1}{\rho}\; dp.
\end{equation}
From eq.~(\ref{firstorder}) (or using any
numerical method of Ref. \cite{EriguchiMuller,Hachisu})
we can see (Fig.~\ref{Isodensity}}) that the shape of the stellar core
is far from spherical or spheroidal one. In case of the rapid differential
rotation the density distribution is toroidal with an off-center maximum. This
``disk-like'' shape (Fig.~\ref{Isodensity}) is sometimes referred to as a
``concave-hamburger''. A common feature of realistic density calculations is a
cusp at the star surface in the polar region. It is a result of the strong
centrifugal force near the rotation axis which is balanced by a rather weak
gravity far from the center.

Using the equilibrium approach we can obtain highly deformed and very flattened
structures, but we can not forget that if they exceed stability limits their structure will be destroyed. Less constrained
rotation, different from cylindrical and rigid one, results in a much more
complicated problem because a physically consistent treatment requires
accounting for the temperature-dependent equation of state and the meridional
circulation in a star. The importance of inclusion of the ``shellular rotation''
and the external pressure is difficult to asses at present. The conclusion is
that the current status of the stellar structure models with rotation does not
allow us to predict firmly the initial state of the collapsing core in the
supernova theory. A non-axisymmetric and ring-like shape of the core is a very
intriguing possibility as it may be partially responsible for violent processes
during the last phases of the presupernova evolution and for the onset of the
core collapse. The growing evidence of both axisymmetric and non-axisymmetric
supernova explosions indicates a significant amount of rotation of the
presupernova core but details are still unclear. It seems thus quite possible
that at least some of the supernova progenitor cores can be properly described
within a framework of the sketched above picture.

}

\subsection{Rotating core collapse}

{

We mentioned already that the ratio $E_{rot}/E_{grav}$ grows as a result of the
core contraction. This happens during subsequent nuclear burning phases and
collapse. When the star exhausts a given nuclear fuel (H, He, C/O, Si) the core
contracts until the next nuclear reactions ignite. After contraction the size of
the core is smaller but the ratio $E_{rot}/E_{grav}$ is bigger. During last
phases of nuclear burning it may reach values as high as 0.04
\cite{HegerandLanger}.

To imagine what may happen in the collapsing and rotating core let's look at the
behavior of a homogeneous and rigidly rotating sphere of a decreasing radius. In
this case:
\begin{equation}
\label{sball}
\frac {E_{rot}}{E_{grav}}\left(R\,\right)
=
\frac
{E^{(0)}_{rot}}{E^{(0)}_{grav}}\,\frac {R_0}{R}
\end{equation}
where quantities with indices ``0'' are just before the beginning of shrinking.
From eq.~(\ref{sball}) we can see that the ratio $E_{rot}/E_{grav}$ grows as
$\sim\!R^{-1}$. From this simple consideration one can infer the existence of
the critical radius for which every stability limit will be exceeded. This
implies impossibility of too big a shrinking even for a moderately fast rotating
body. The process of shrinking will be stopped by the body breakup or another
catastrophe.

During the collapse of the core its radius, $R$, is decreasing very fast. As we
mentioned, in calculations of \cite{HegerandLanger} the iron cores with
$E_{rot}/E_{grav}=0.01 \dots 0.04$ and with the initial radii $R_0 \approx
2000$km are found. The final radius is $R \approx 10$km -- a typical value of
the neutron star radius. Actually, just after the collapse the radius of
a newborn protoneutron star is $R \approx 60$km, and after about one second it
shrinks to $\approx 20$km. Finally, after tens of seconds the radius sets at
$R\approx10$km as a result of deleptonization. The dynamical instability limit
of 0.27 for Mac Laurin spheroid is exceeded if the Fe core radius shrinks to $80
\dots 300$km, respectively, for the initial ratio $E_{rot}/E_{grav}=0.01 \dots
0.04$. This indicates the possibility of a violent hydrodynamical instability.
Because the shock front is born at approximately the same radius and at the same
time, only very detailed dynamical simulations can give the actual behavior of
the collapsing core.

Recently, results of full three-dimensional simulations with no symmetry of the
collapsing core assumed have been published \cite{Rampp}. In this paper a
simplified equation of state and initial models of \cite{ZwergerandMuller} were
used. This work focused on the gravitational radiation emission from
the core. The core gets rid of the extra angular momentum in the form of the
spiral arms or a small ``satellite'' as can be seen (see footnote \ref{anims}
on page \pageref{anims} for WWW address) in the results of Rampp \cite{Rampp}.
Possible future detections of gravitational waves correlated with SN explosions
by the next generation detectors  (LISA, LIGO) will be an ultimate proof of the
strongly asymmetric processes in the supernova engine.

Detailed numerical calculations of the supernova explosion with inclusion of the
neutrino processes have been reported in \cite{FryerandHeger,YamadaandSato} for
an axisymmetric case with the equatorial symmetry assumed. However, the obtained
results are rather ambiguous (see next section).

}

\section{Rotation and extended classification scheme \label{LastSec}}

{

In previous sections we have discussed some problems encountered when
one attempts to include rotation in the core-collapse supernova theory. A
question thus arises is such an extension of the theory necessary from the
observational point of view? The answer is definitively affirmative. Presently
available astronomical data relevant to the supernova phenomenon suggest
strong non-sphericity of the explosion. The most important pieces of evidence
include velocities of pulsars sometimes correlated with an apparent deformation
of remnants and the asymmetry of explosion deduced from the measured
polarization. Also, some events appear to be superluminous under assumption of
the spherical symmetry. Finally, the very recent identification of the supernova
component in the light curve of GRB011121 provides a compelling argument in
favour of the core rotation as discussed below.

Let us explain how the core rotation can account for the those features of
the data that can not be understood in the standard supernova scenario.

\begin{figure}
\centering
\includegraphics[height=0.9\textwidth,angle=270]{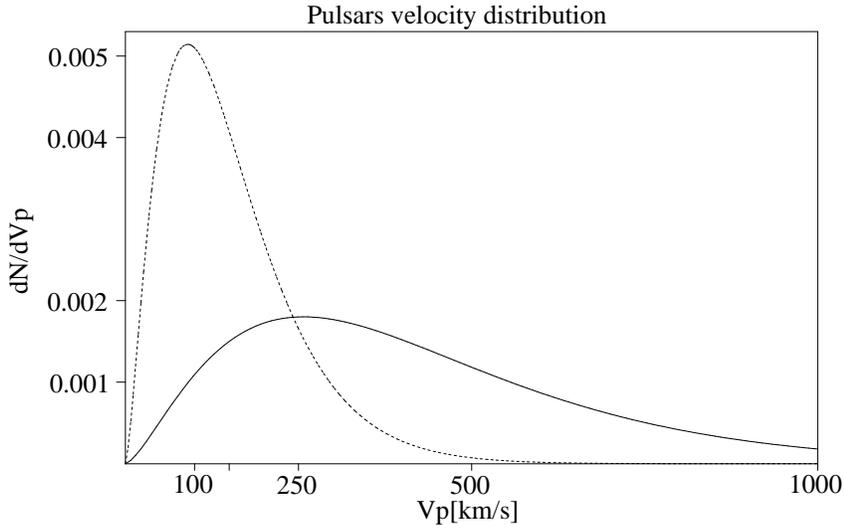}
\caption{Distributions of pulsar velocities. The solid line corresponds to the
observed velocities and the dashed line to natal velocities obtained by
subtracting  binary effects. Both solid and dashed curves are described by a
$\chi^2$-like
distribution, $f(v_p) = a\,v_p^{3/2}\,exp(-v_p/b)$, adopted from
\cite{Vanbeveren} with  $a=1.96\times10^{-6}\; b=514/3$ and
$a=2.7\times10^{-5} \; b=60$, respectively.}
\label{Vp}
\end{figure}

Pulsars have much higher velocities than the average velocity of stars in the
disk of the Galaxy. The highest observed speed exceeds 2300 km/s
\cite{Glendening}. The observed pulsar speed \cite{Vanbeveren} is generally a
sum of two components:
\begin{itemize}
\item a ``kick" imparted to the nascent neutron
star during the supernova explosion
\item an inherited velocity due to progenitor revolution in a binary system --
when SN occurs in a binary (what happens quite often, probably most SN are in
binaries)
\end{itemize}
If the explosion disrupts a binary system we can see a runaway pulsar. The
existence of binary pulsars tells us that some of systems survive two SN events.
This also generally requires some momentum transfer to reduce the orbital
velocity because of dramatic mass loss during the explosion.
The velocity component due to the pulsar's birth in a binary system can be
subtracted statistically from the distribution of pulsar velocities. We then
obtain the distribution of kick velocities directly related to the SN explosion
(Fig.~\ref{Vp}). These velocities are still large enough to challenge
the results of spherically symmetric supernova models. Proposed explanations
of such large kicks include \cite{Lai}:
\begin{itemize}
\item hydrodynamically driven kicks
(fluctuations in {\it a non-rotating core collapse})
\item asymmetric neutrino emission
\item electromagnetic post-natal ``rocket effect''
\end{itemize}
All these models disregard any dynamically significant rotation. As a
kinematical effect, the rotation adds some stochastic component to such
processes as the convective motion or the neutrino emission.

Although kicks in rapidly rotating supernova models which lead to
non-axisymmetric deformations have not been studied systematically, the
asymmetric disruption of a rapidly rotating core can be seen in some
calculations. Results presented in \cite{Rampp} show how a non-zero component of
the velocity perpendicular to the rotation axis arises in simulations of a
formation of PNS\footnote{http://www.mpa-garching.mpg.de/\~{}wfk/MOVIES/
\label{anims}}.

It is in principle possible to distinguish between kicks produced with a
different amount of rotation if the relative orientation of the spin and the
direction of motion is known. When the rotation is slow with respect to the kick
mechanism there is no correlation between the rotation axis and the kick
direction. When the rotation is dynamically unimportant but faster than the kick
mechanism, the angular averaging effect leads to the parallel orientation.
For very fast rotation, leading to the asymmetric disruption of the core, we
expect the rotation axis to be perpendicular to the momentum.
In all three cases the relative orientation of the current velocity vector and
the rotation axis can be found. The analysis of the binary neutron star system
B1913+16 \cite{Wex} gives the result that the angle between the kick direction
and the orbital plane is less than $5^{\circ}-10^{\circ}$. If the orbital
momentum and the spin were aligned before the explosion, as it was assumed in
the cited article, we would get an almost perpendicular orientation of the kick
and the PNS rotation axis. The study of Vela and Crab pulsars \cite{Lai} based
on observed jets shows an apparent alignment on the sky of the velocity and the
rotation axis.

Statistical analysis of other 28 pulsars shows no correlation between their
motion and rotation \cite{Deshpande}. Ambiguity of these results is due to
difficulties in the analysis itself, simplifying assumptions and poor quality
of the data. We only note that perpendicularity of the rotation axis and the
kick, favored by the asymmetrical disruption of a rapidly rotating core in the
SN explosion, is not disproved from the observational point of view.

We also note that the fast rotation of pulsars is a strong evidence in favour of
a huge angular momentum of the central part of exploding star.
Rotation of the neutron star is the immediate consequence of the presupernova
core rotation. There is no need to invoke any other mechanism, such as \eg an
off-center kick, to account for fast rotation of pulsars, in contrast to stellar
models with very slow rotation of the presupernova core.

\subsection{GRB - SN connection}

Next argument in favour of strong rotation in supernovae comes from observations
of superluminous events like SN1998bw -- well known from its controversial
coincidence with the gamma ray-burst GRB980425. Spectral analysis classified it
as type Ic, but the absolute magnitude was close to that of type Ia
\cite{Kulkarni,Galama}. A spherically symmetric modelling has given an enormous
explosion energy of 20--50~foe, the ejected mass in the range $12-15 M_{\odot}$
and the radioactive nickel amount as big as $0.5 - 0.8M_{\odot}$ \cite{Iwamoto}.
Listed values are considered impossible in the standard supernova scenario. That
is why these events are often called hypernovae\footnote{This in not a hypernova
in the sense of the theoretical GRB models, but the possibility of a real
connection between the two is not excluded and is a topic of current research}.
As it has been shown in \cite{Holfich} observed results could be well understood
if the prolate asymmetry of the expansion, with the axis ratio 2:1, is assumed.
We note, that the prolate expansion velocity produces oblate iso-density
contours. This requires the standard explosion energy of 2~foe if observed
$60^{\circ}$ above the plane of symmetry. The ejected mass and the $^{56}$Ni
amount are $2M_{\odot}$ and $0.2M_{\odot}$, respectively -- typical values for a
luminous core-collapse supernova. Polarization of light, which is more directly
related to the asymmetry of expansion, is often present during hypernova events,
again indicating a non-spherical explosion. Analysis of this sort, based on the
rescaling of a spherical model tells us nothing about physical processes
responsible for the asymmetry, but the rotation of the collapsing core is one of
the most probable reasons.

Very recently, observations of a SN event in the light curve of GRB011121, have
been reported \cite{Bloom}. Joint campaign of optical and X-ray observations,
also from space, succeeded to firmly establish the presence of a supernova in
the GRB011121 place, when the afterglow has declined. The supernova is of type
Ic. Collected observations of GRB011121 fit nicely to the collapsar model of
MacFadyen and Woosley \cite{collapsar}. In this model, rotation of the
progenitor's core plays a crucial role and the rotation axis determines the
direction of jets which produce observed gamma rays. Despite the formation of a
black hole from the collapsing core, matter rich in $^{56}$Ni is expelled along
the rotation axis and its emission forms a typical supernova light curve
\cite{collapsar}.

\subsection{Including rotation into the classification scheme}

\begin{figure}[t]
\begin{center}
\includegraphics[width=\textwidth]{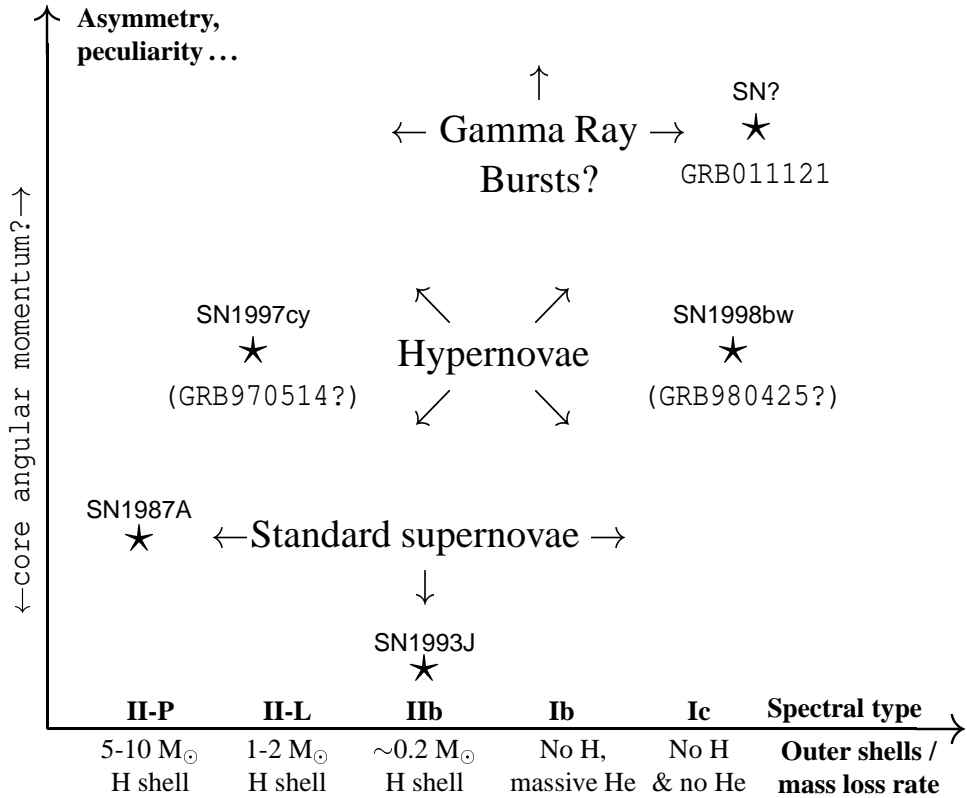}
\end{center}
\caption{Two-dimensional classification of supernovae and
some GRB events. The vertical axis represents a property which is can serve as
a measure of the core angular momentum}
\label{2D}
\end{figure}

We feel encouraged by the above analysis to unify the phenomena of supernovae,
hypernovae and some GRB's in a two-dimensional classification scheme which
will expand that shown in Fig.~\ref{1D}. As a second dimension we include some
measure of rotation of the progenitor's core (\eg its angular momentum). We
conjecture that classical supernovae correspond to some relatively low amount
of angular momentum, which is just enough to produce a neutron star with
proper natal momentum. Here we assume that the explosion itself may be treated
as a strong argument in favour of a sufficiently high rotation of the
presupernova core. The simulations have shown the failure of the idea of
explosion by the prompt mechanism. The delayed mechanism improved the situation
but it is not clear at all if it really works. Indeed, some of the models still
fail to produce SN events in spite of inclusion of the neutrino processes.

When the core angular momentum if bigger one expects that some part of the
collapsing core is locked for a while in a sort of "fizzler" and is accreted
back onto the inner part. Depending on the falling back mass, a neutron star can
form, or further collapse to black hole can proceed. As the fizzler possess
large angular momentum, very fast rotating neutron star can form, if the
accreted mass does not exceed some fraction of the solar mass. In case the mass
is much higher, of a couple of solar masses, as in numerical simulations happens
for heavier iron cores formed in massive helium stars (of say 10 solar masses),
a Kerr black hole finally forms. A transient accretion disk feeds angular
momentum to the black hole and cools emitting strong neutrino flux. One can
conjecture, that hypernovae and GRB's are displays corresponding to such a way
of the core collapse. Positions of various phenomena in the two-dimensional
classification scheme are shown in Fig.~\ref{2D}.

\subsection{Discussion}

If simulations of the rotating supernova fail in the sense that the energy of
radiation and ejecta  is similar to that obtained in the simulations without
rotation which give a sufficient explosion power we have to accept standard
scenario. In the opposite situation we have an indirect argument in favour of
rotation of the Fe core. The reasoning presented above holds provided the
results of simulations of such a complex problem as the supernova engine are
reliable. One should notice that some serious difficulties are reflected in the
published results of simulations with rotation. Yamada and Sato
\cite{YamadaandSato} found stronger explosion but recently Fryer and Heger
\cite{FryerandHeger} obtained decrease of explosion energy. }

Finally we note that the evidence of strong asphericity in supernova explosions
is being established now. Besides the arguments mentioned in the previous
section, some very suggestive images of supernova remnants are available. One of
them is Puppis A with the asymmetry apparently correlated with the movement of a
young neutron star \cite{PuppisA}. Pictures like this are very suggestive and
tell us that pulsar kicks are natal and produced by the supernova engine. A lot
of processes may be responsible for this, however rotation is the most common
feature of astrophysical objects and it is reasonable to examine its possible
effects first. The knowledge relevant to a wide range of problems related to
rotation, such as the initial pre-collapse state and the behavior of the
contracting core which exceeds the rotation stability limits, is still
unsatisfactory. Advances in the supernova rotation research will hopefully lead
to a better understanding of many problems of supernova physics, such as the
mechanism of pulsar kicks, the GRB-SN connection, the emission of gravitational
waves and the formation of Kerr black holes.

The authors are grateful to W.~J.~\'Swiatecki for critical reading of the manuscript and for claryfing
 discussion on stability of Jacobi ellipsoids \cite{SwiateckiLBL}.

\end{document}